\documentclass[runningheads]{llncs}
\usepackage[T1]{fontenc}

\usepackage{cite}
\usepackage{amsmath,amssymb,amsfonts}
\usepackage{graphicx}
\usepackage{textcomp}
\usepackage{xcolor}
\usepackage{todonotes}

\usepackage{macros}
\usepackage[bookmarks=false]{hyperref}
\usepackage{relsize}
\usepackage{multirow}
\usepackage{colortbl}
\usepackage{subcaption}
\usepackage{adjustbox}
\usepackage{relsize}
\usepackage{shuffle}
\usepackage{fontawesome5}
\usepackage[utf8]{inputenc}
\usepackage{marginnote}
\usepackage{enumitem}

\usepackage[ruled,vlined,linesnumbered]{algorithm2e}

\expandafter\def\expandafter\normalsize\expandafter{%
    \normalsize%
    \setlength\abovedisplayskip{3pt}%
    \setlength\belowdisplayskip{3pt}%
    \setlength\abovedisplayshortskip{3pt}%
    \setlength\belowdisplayshortskip{3pt}%
}

\AtEndEnvironment{figure}{\vspace{-0.3cm}}
\AtEndEnvironment{figure*}{\vspace{-0.3cm}}

\makeatletter
\let\oldnl\nl
\newcommand{\nonl}{\renewcommand{\nl}{\let\nl\oldnl}}
\makeatother

\setlist{topsep=0pt, partopsep=3pt, parsep=0pt, itemsep=2pt}

\makeatletter
\renewcommand{\section}{\@startsection{section}{1}{0pt}%
  {3ex plus 1ex minus .2ex}
  {1.5ex plus .2ex}
  {\normalfont\Large\bfseries}} 
\makeatother

\makeatletter
\renewcommand{\subsection}{\@startsection{subsection}{2}{0pt}%
  {1.5ex plus 1ex minus .2ex}
  {0.8ex plus .2ex}
  {\normalfont\normalsize\bfseries}}  
\makeatother

\makeatletter
\renewcommand{\subsubsection}{\@startsection{subsubsection}{3}{0pt}%
  {1.5ex plus 1ex minus .2ex}%
  {0.8ex plus .2ex}%
  {\normalfont\normalsize\bfseries}}
\makeatother

\begin{document}

\title{Revealing Inherent Concurrency in Event Data: A Partial Order Approach to Process Discovery\thanks{The Version of Record of this contribution will be published in the proceedings of the 1st International Workshop on Stochastics, Uncertainty and Non-Determinism in Process Mining (SUN-PM). This preprint has not undergone peer review or any post-submission improvements or corrections.}}
\titlerunning{Revealing Inherent Concurrency in Event Data}

\author{Humam Kourani\inst{1,2}\orcidID{0000-0003-2375-2152} \and
Gyunam Park\inst{1}\orcidID{0000-0001-9394-6513} \and
Wil M.P. van der Aalst\inst{1,2}\orcidID{0000-0002-0955-6940}}
\authorrunning{H. Kourani et al.}  

\institute{Fraunhofer FIT, Schloss Birlinghoven, 53757 Sankt Augustin, Germany\\
\email{\{humam.kourani,gyunam.park,wil.van.der.aalst\}@fit.fraunhofer.de} \and
RWTH Aachen University, Ahornstraße 55, 52074 Aachen, Germany}

\maketitle

\begin{abstract} 
Process discovery algorithms traditionally linearize events, failing to capture the inherent concurrency of real-world processes. While some techniques can handle partially ordered data, they often struggle with scalability on large event logs. We introduce a novel, scalable algorithm that directly leverages partial orders in process discovery. Our approach derives partially ordered traces from event data and aggregates them into a sound-by-construction, perfectly fitting process model. Our hierarchical algorithm preserves inherent concurrency while systematically abstracting exclusive choices and loop patterns, enhancing model compactness and precision. We have implemented our technique and demonstrated its applicability on complex real-life event logs. Our work contributes a scalable solution for a more faithful representation of process behavior, especially when concurrency is prevalent in event data.


\end{abstract}


\section{Introduction}

Process discovery, a fundamental pillar of process mining, focuses on automatically generating a process model from such an event log, providing a visual and analytical representation of the underlying process behavior. 

Traditional process discovery techniques assume totally ordered event data, where each process instance is a strict activity sequence (e.g., $\langle a, b, c\rangle$). This clashes with the reality of complex processes where activities frequently execute concurrently or overlap in time. While information systems may record interval data (e.g., start and completion timestamps), conventional methods often discard this by selecting a single timestamp, losing information about true execution semantics. Even with atomic events, events might share the exact same timestamp, naturally highlighting their lack of strict sequential dependency. Furthermore, recorded timestamps may be unreliable, suffer from varying granularities, or be recorded at a level too fine-grained to reveal meaningful process structure \cite{DBLP:journals/jdiq/HofstedeKMAFSWCWGMS23}. Discrepancies between actual event occurrence and recording times are common, particularly in domains like healthcare where manual data entry delays are frequent \cite{bc469e8cb3744634a6ab0b2149802b9a}. In such cases, an abstraction of timestamps to a coarser level (e.g., hour, day, or even coarser business-defined periods) becomes necessary.

\begin{figure}[!t]

        \centering
        \includegraphics[width=0.18\linewidth]{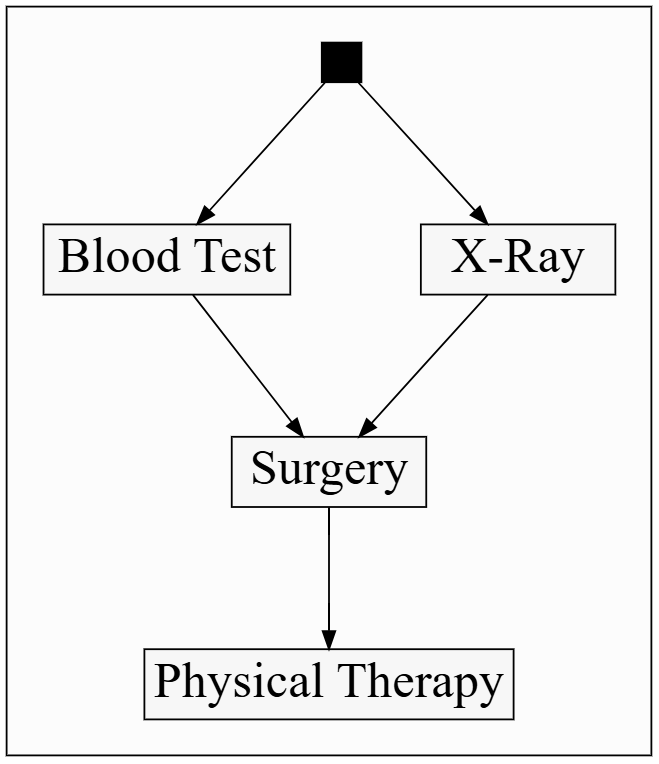}
        \includegraphics[width=0.18\linewidth]{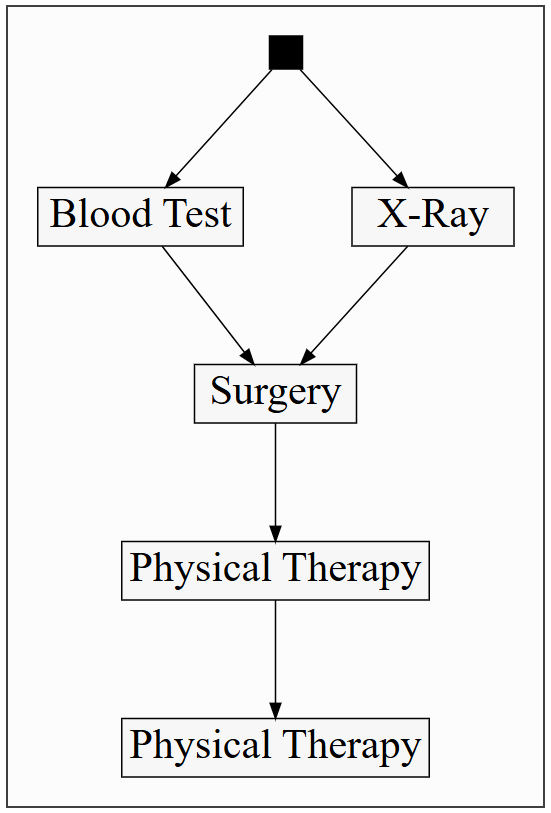}
        \includegraphics[width=0.18\linewidth]{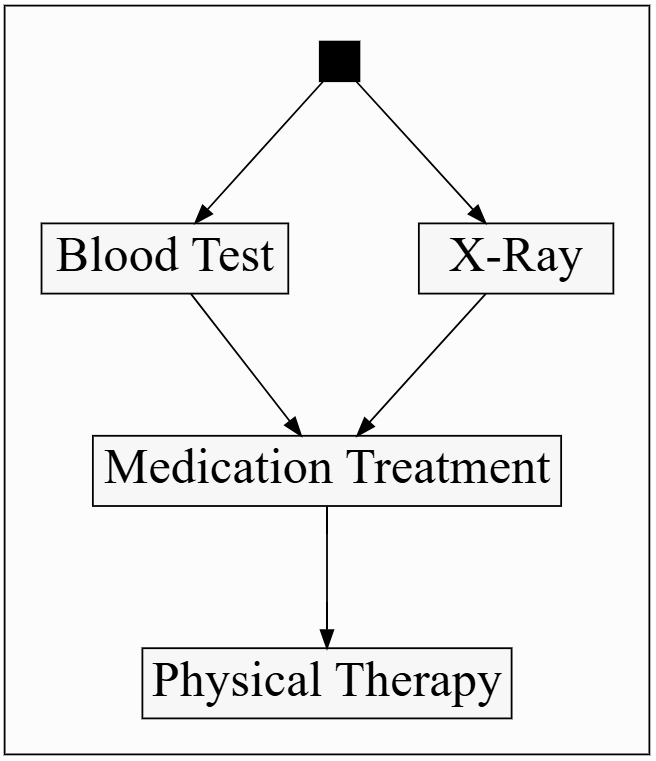}
        \includegraphics[width=0.18\linewidth]{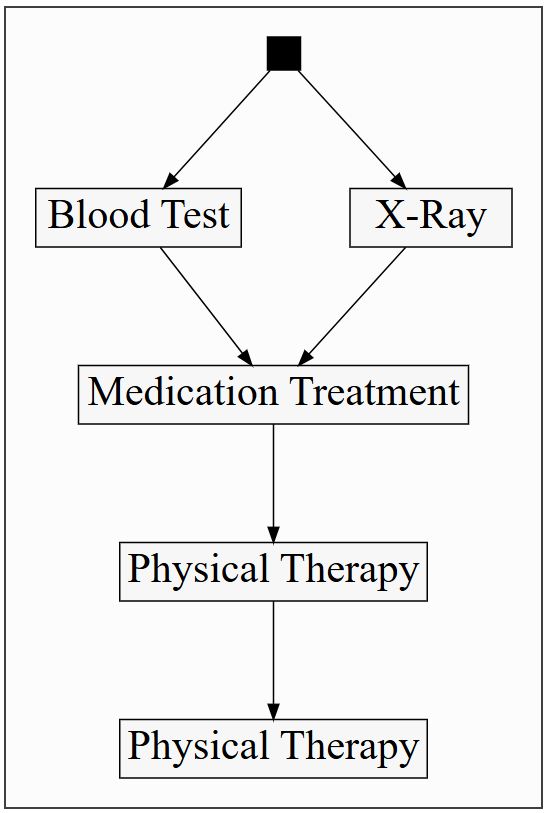}
        \caption{Trace variants for the hospital process, represented as four partial orders.}
        \label{fig:ex:medical:variants}

\end{figure}

Consider a simplified hospital patient journey. A nurse takes a blood sample but logs it hours later (e.g., at the end of the shift). Later that evening, the patient undergoes an X-ray, but only the date of this procedure is logged (00:00 timestamp). 
In the next days, a choice is made: either a surgery is needed or a medication treatment is sufficient. Towards the end of the process, one or possibly two courses of physical therapy are prescribed. In this hospital scenario, imposing a total order based on raw timestamps would be misleading. Abstracting timestamps to the `day' level is necessary, revealing underlying partial orders, as shown by the four trace variants in \autoref{fig:ex:medical:variants}. This example underscores the necessity for discovery techniques for partially ordered data. Traditional approaches that assume an arbitrary order for events with identical timestamps introduce fictitious sequential dependencies that misrepresent the actual process logic.

This paper proposes a novel process discovery approach that leverages the Partially Ordered Workflow Language (POWL) \cite{DBLP:conf/bpm/KouraniZ23}. POWL is a hierarchical process modeling language where submodels are combined to form larger ones, either as partial orders or using control-flow operators: \xor\ for exclusive choice and \Loop\ for repeated behavior. Crucially, POWL guarantees soundness by construction, meaning that POWL models can be translated into Workflow nets that inherently free from common structural issues such as deadlocks, a highly desirable trait for process discovery. Furthermore, POWL's design, which inherently supports partial orders as a fundamental modeling construct, makes it an ideal process modeling language for discovering processes from partially ordered data. Our methodology first derives partial orders from event data. Our algorithm then merges them into a general, encompassing partial order. It also incorporates additional steps to detect common control-flow patterns: mutually exclusive activities are grouped using \xor, and repeatable blocks with \Loop. This results in a POWL model that can more accurately capture real-life process behaviors, especially their inherent concurrency.



\section{Related Work}\label{sec:related_work}
While POWL has been used for process discovery in \cite{DBLP:journals/is/KouraniZSA25, DBLP:conf/icpm/KouraniSA23}, these existing techniques follow traditional discovery paradigms that operate on the assumption of totally ordered traces. The investigation of process mining techniques that handle inherent concurrency in event data has become a significant and growing research area, as surveyed in \cite{DBLP:journals/kais/LeemansZL23}. In \cite{DBLP:conf/bpm/LeemansFA15}, the authors introduce an Inductive Miner variant that leverages lifecycle transition data to differentiate concurrency from sequential interleaving. Similarly, the Split Miner is refined in \cite{DBLP:conf/icpm/AugustoDR20} to detect overlapping execution intervals when start/end timestamps are available.  In \cite{DBLP:conf/apn/DumasG15}, the authors advocate for using prime event structures as a unified representation for both process models and event logs, highlighting their natural ability to capture concurrency. The Prime Miner \cite{DBLP:conf/icpm/Bergenthum19} leverages lifecycle data to discover prime event structures from event logs and finally synthesize a Petri net via region-based techniques. 

The Multi-Phase Miner \cite{van2005multi} aggregates instance graphs into higher-level models. Further exploring aggregation, the authors in \cite{DBLP:journals/topnoc/DongenDA12} detail algorithms for constructing workflow nets from sets of partially-ordered causal runs. The eST\textsuperscript{2} Miner \cite{folz2025est} combines a replay-based discovery algorithm with Petri net synthesis from partial orders. The ILP\textsuperscript{2} Miner \cite{DBLP:conf/apn/FolzWeinsteinBDK23} uses Integer Linear Programming for discovery from partially ordered event logs. The Zebra Miner \cite{kovavr2024exploratory} employs an incremental region-based technique to synthesize Petri nets from partial orders. Another synthesis approach uses Petri net unfoldings, where an expert-provided independence relation is used to construct partial orders \cite{DBLP:conf/atva/LeonRCHH15}. Finally, the authors in \cite{DBLP:conf/apn/Nombre11a} explore the synthesis of Petri nets by folding partially ordered runs. 

Existing techniques, particularly those relying on computationally intensive synthesis like region-based mining or integer linear programming, often face scalability challenges on large, complex event logs \cite{DBLP:journals/kais/LeemansZL23}. Our hierarchical and aggregative method is designed for efficiency, making it applicable to real-world datasets.

\section{Preliminaries}\label{sec:preliminaries}

A \emph{multiset} over a set \( X \), denoted \( M \in \bag(X) \), allows for multiple occurrences of elements from \( X \). 
For a set \( X \), a \emph{partition} of \( X \) of size \( n \geq 1 \) is a set of subsets \( P = \{X_1, \dots, X_n\} \) such that \( X = X_1 \cup \dots \cup X_n \), \( X_i \neq \emptyset \) for all \( 1 \leq i \leq n \), and \( X_i \cap X_j = \emptyset \) for all \( 1 \leq i < j \leq n \). 

A binary relation \( \po \subseteq X \times X \) over a set \( X \) indicates relationships between pairs of elements. We write \( x_1 \po x_2 \) for \( (x_1, x_2) \in \po \). The \emph{transitive closure} of $\po$ is defined as $\closure\po = \{(x, y) \mid \exists_{x_1, \dots, x_n \in X} \ x = x_1 \ \wedge \ y = x_n \ \wedge \forall_{1\leq i<n} x_i \po x_{i+1}\}$.

A \emph{strict partial order} (or simply \emph{partial order}) $\po$ over a set $X$ is an irreflexive ($x \notpo x$ for all $x \in X$) and transitive ($x_1 \po x_2 \wedge x_2 \po x_3 \Rightarrow x_1 \po x_3$) binary relation. These properties imply asymmetry ($x_1 \po x_2 \Rightarrow x_2 \notpo x_1$). Given a set \( X \), we use \( \mathcal{O}(X) \) to denote the set of all partial orders over \( X \).

Let $\ActUniverse$ be the universe of activity labels. An \emph{Event Log} is a collection of events. An event $e$ has at least the following attributes: a case identifier, an activity label $\lab(e) \in \ActUniverse$, and a timestamp. Optionally, an event may have a lifecycle attribute to distinguish different phases of the activity execution (e.g., \texttt{start}, \texttt{complete}, \texttt{suspend}, and \texttt{resume}). Events that share the same case identifier represent the execution of a single process instance (often referred to as a \emph{trace} or \emph{case}).

We use $\tau \notin \ActUniverse$ to denote a silent activity, which represents an unobservable action. Let $\TraUniverse$ be the universe of transitions. Each transition $t \in \TraUniverse$ has an associated activity label $\lab(t) \in \ActUniverse \cup \{\tau\}$. Transitions having the same label represent multiple instances of an activity. In this paper, we conceptualize a transition $t$ as a pair $(a, k)$ where $a = \lab(t)$ is its label and $k = idx(t)$ is an index that distinguishes it from other transitions with the same label within the same process instance. For example, we use $t_1 = (a,1) \in \TraUniverse$ and $t_2 = (a,2) \in \TraUniverse$ to represent two distinct instances of an activity $a \in \ActUniverse$.

The Partially Ordered Workflow Language (POWL) \cite{DBLP:conf/bpm/KouraniZ23} is a hierarchical process modeling language. POWL models can be combined to form larger ones using a partial order, the choice operator (\xor), or the loop operator (\Loop).

\begin{definition}[POWL Model]\label{def:powl}
POWL models are defined recursively:
\begin{itemize}
    \item Any transition $t \in \TraUniverse$ is a POWL model.
    \item Let $P = \{ \powl_1, \dots, \powl_n \}$ be a set of $n \geq 2$ POWL models. Then:
    \begin{itemize}
        \item $\xor(P)$ is a POWL model.
        \item $\Loop(\powl_1, \powl_2)$ is a POWL model.
        \item A partial order $\po \in \mathcal{O}(P)$ is a POWL model. 
    \end{itemize}
\end{itemize}
\end{definition}

The semantics of a POWL model are given by the set of activity sequences it can produce. For \xor, only one of the submodels is executed. For \Loop, the first submodel (do-part) is executed, then zero or more repetitions of the second submodel (redo-part) followed by the do-part are executed. In a partial order, all submodels are executed while respecting the given execution order.

Two POWL models are considered equivalent if they are of the same type and their children are recursively equivalent. For atomic transitions, equivalence is based on having the same activity label.

\begin{definition}[Equivalent POWL Models]\label{def:equivalent}
Let $\powl$ and $\powl'$ be two POWL models. The two models are equivalent, denoted as $\powl \eq \powl'$, if one of the following conditions is met:
\begin{itemize}
    \item $\powl \in \TraUniverse \ \wedge \ \powl' \in \TraUniverse \ \wedge \ \lab(\powl) = \lab(\powl')$.
    \item Two sets of $n \geq 2$ POWL models $P = \{\powl_1, \dots, \powl_n\}$ and $P' = \{\powl'_{1}, \dots, \powl'_{n}\}$ 
    exist such that $\powl_i \eq \powl'_{i}$ for any $1\leq i \leq n$ and one of the following conditions is met:
        \begin{itemize}
            \item $\powl = \xor(P)$ and $\powl' = \xor(P')$.
            \item $n=2$, $\powl = \Loop(\powl_1, \powl_2)$, and $\powl' = \Loop(\powl'_{1}, \powl'_{2})$.
            \item $\powl = \po \in \mathcal{O}(P)$ and $\powl' = \po' \in \mathcal{O}(P')$ such that $\powl_i \po \powl_j \Leftrightarrow \powl'_{i} \po' \powl'_j \text{ for all }\allowbreak i, j \in \{1, \dots, n\}$.
        \end{itemize}
\end{itemize}
\end{definition}

We use $\mathcal{L}(\powl)$ to denote the set of all activity labels present within the hierarchical structure of a POWL model $\powl$. For example, if $\powl = \Loop(\xor((a, 1), (b, 1)),\allowbreak (a, 2))$, then $\mathcal{L}(\powl) = \{a, b\}$.

\section{Deriving Partial Orders from Event Data}\label{sec:deriving_pots}
To enable our process discovery approach, raw event data must first be transformed. This section details the critical preparatory stage: the systematic conversion of a standard event log into a collection of partial orders. 


An \emph{interval event log} $L_{int}$ is a collection of \emph{interval events}, where each interval event $e$ represents a single, complete execution of an activity. Formally, an interval event can be represented as a tuple $e = (a, c, st, et)$, where $a \in \ActUniverse$ is the activity label, $c$ is the case identifier, $st$ is the start timestamp, and $et$ is the completion timestamp, with $st \leq et$. We use $\mathcal{C}(L_{int})$ to denote the set of all case identifiers in an interval event log, i.e., $\mathcal{C}(L_{int}) = \{c \mid (a, c, st, et) \in L_{int}\}$.

The precise method of deriving an interval event log $L_{int}$ from a raw event log $L$ can vary and is highly configurable. In our work, we employ a method that matches `start' and `complete' lifecycle transitions for the same activity label within each case on a First-In, First-Out (FIFO) basis to form interval events. If lifecycle information is absent or any `complete' events could not be matched via the FIFO process, then atomic interval events are created (i.e., with identical start and end timestamps). This approach, similarly to the one in \cite{DBLP:conf/bpm/LeemansFA15}, enables the direct application of the downstream discovery algorithm to any event log, even when lifecycle information is incomplete.

Once $L_{int}$ is obtained, we can construct a partial order to represent each process instance.

\begin{definition}[Partially Ordered Trace]\label{def:pot_from_interval_trace_final}
Let $L_{int}$ be an interval event log and $c \in \mathcal{C}(L_{int})$ be a case identifier in $L_{int}$. Let $e_1 = (a_1, c, st_1, et_1), \dots, e_n = (a_n, c, st_n, et_n) \in L_{int}$ be all interval events that have $c$ as their case identifier in $L_{int}$. The corresponding \emph{Partially Ordered Trace (POT)} $P_c$ is a pair $(V_c, \po_c)$, where:
\begin{itemize}
    \item $V_c = \{t_1, \dots, t_n\}$ the a set of transitions, where each transition corresponds to an interval event, i.e., $\lab(t_i) = a_i$ for all $1 \leq i \leq n$.
    \item $\po_c \subseteq V_c \times V_c$ is a binary relation defined as follows for any $i, j \in \{1, \dots, n\}$: $t_i \po_c t_j \iff et_i < st_j$.
\end{itemize}
\end{definition}

For consistency, we always use positive integers (1, 2, ...) as indexes for transitions sharing the same activity label within a POT, ordered by their start and end timestamps. For example, if two interval events within a case $c$ share the same label $a$, the one with the earlier start timestamp is mapped into $(a, 1)$, and the one starting later is mapped into $(a, 2)$. 

The relation $\po_c$ in \autoref{def:pot_from_interval_trace_final} captures the temporal precedence between activity instances: $t_i \po t_j$ if $e_i$ completes strictly before $e_j$ starts. This definition naturally handles concurrency, as activity instances whose execution intervals overlap will not have a precedence relation between them. It is straightforward to show that $\po_c$ is a strict partial order. The use of transitions ensures that even if multiple instances share the same activity label, they are treated as distinct nodes in $V_c$, allowing $\po_c$ to be well-defined as a strict partial order based on the individual start and completion times.

Given an interval event log $L_{int}$, the multiset of all POTs derived from $L_{int}$ serves as the direct input to our POWL discovery algorithm.

\begin{definition}[Partially Ordered Event Log]
Let $L_{int}$ be an interval event log. The partially ordered event log corresponding to $L_{int}$ is the multiset of POTs defined as follows: $L_{PO} = [ P_c \mid c \in \mathcal{C}(L_{int})]$.
\end{definition}




\section{POWL Discovery from Partial Orders}\label{sec:methodology}
This section presents our algorithm for discovering a POWL process model from a multiset of POTs. The core of our algorithm is the aggregation of the input partial orders into a single, encompassing partial order that captures the overall process structure. The algorithm also incorporates additional steps that aim to identify and abstract exclusive choice, repeated, and co-occurring patterns.

\subsection{Notation and Basic Operations}

Let $M$ be a multiset of partial orders. The set of all unique nodes present across all partial orders in $M$ is denoted by $\mathcal{V}(M) = \bigcup_{(V, \po) \in M} V$.

During recursive discovery, it's often necessary to isolate the behavior related to a specific subset of nodes. For instance, when an XOR is detected, we need to independently discover a model for each of its branches. The \emph{partial order projection} operation enables this by creating a new multiset of partial orders that only contain elements from a specified subset of nodes.


\begin{definition}[Partial Order Projection]\label{def:proj}
Let $M$ be a multiset of partial orders and let $V_{sub} \subseteq \mathcal{V}(M)$ be a subset of nodes in $M$. Then the projection of $M$ on $V_{sub}$ is defined as follows:
\[
M\proj_{V_{sub}} \ = [(V', \po \cap (V' \times V')) \mid (V, \po) \in M \]
\[\wedge \ V' = V \cap V_{sub} \ \wedge \ V' \neq \emptyset].
\]
\end{definition}

Once a structural pattern (e.g., an XOR) involving a set of nodes $V_{sub}$ is identified and a corresponding POWL submodel $\powl$ is created, the algorithm replaces $V_{sub}$ with $\powl$ in the current multiset of partial orders $M$. The node substitution must ensure that the new node $\powl$ correctly inherits the precedence relations of the set $V_{sub}$ it replaces.

\begin{definition}[Node Substitution in Partial Orders]\label{def:subst}
Let $M$ be a multiset of partial orders and let $V_{sub} \subseteq \mathcal{V}(M)$ be a subset of nodes in $M$. Let $\powl \notin \mathcal{V}(M)$ be a new node not present in $M$. The substitution of $V_{sub}$ by $\powl$ in $M$ is defined as 
\[
M_{V_{sub} \rightarrow \powl} = [(V', \po') \mid (V, \po) \in M]
\] 
where
\[
V' = 
\begin{cases}
    V & \text{if } V \cap V_{sub} = \emptyset, \\
    (V \setminus V_{sub}) \cup \{\powl\} & \text{otherwise.}
 \end{cases}
\]
and 
\[
\po' = \{(s, t) \mid s\po t \ \wedge \ s \notin V_{sub} \ \wedge \ t \notin V_{sub}\}
\]
\[
\cup \ \{(\powl, t) \mid t\in V \ \wedge \ V\cap V_{sub} \neq \emptyset \ \wedge \ \forall_{s\in V\cap V_{sub}} \ s\po t\}
\]
\[
\cup \ \{(s, \powl) \mid s\in V \ \wedge \ V\cap V_{sub} \neq \emptyset \ \wedge \ \forall_{t\in V\cap V_{sub}} \ s\po t\}.\]
\end{definition}

\subsection{Mining for Exclusive Choices}\label{sec:xor_identification} 
Our discovery algorithm identifies candidates for exclusive choice (XOR) constructs by analyzing the co-occurrence patterns of activities in the input POTs. If certain activities (or sets of activities) are never observed to occur together within the same process instance (trace), they are likely to be mutually exclusive alternatives. To formalize this, we first define a conflict relation between individual activity labels based on their absence of co-occurrence.

\begin{definition}[Activity Conflict Relation]\label{def:conflict}
Let $M$ be a multiset of partial orders, and let $a$ and $b$ be two activity labels present in $M$, i.e., $a, b \in \bigcup_{\powl \in \mathcal{V}(M)} \mathcal{L}(\powl)$. We say that $a$ and $b$ are conflicting in $M$, denoted as $a\#_{M}b$, if and only if:
\[
\nexists_{(V, \po) \in M;\ \powl, \powl' \in V} \ a \in \mathcal{L}(\powl) \ \wedge \ b \in \mathcal{L}(\powl').
\]
\end{definition}

An \emph{activity conflict group} extends this concept to a partitioning of a set of activities, where each part (a set of activity labels) is mutually exclusive with every other part in the partition.

\begin{definition}[Activity Conflict Group]\label{def:conflictgroup}
Let $M$ be a multiset of partial orders. Let $A \subseteq  \bigcup_{\powl \in \mathcal{V}(M)} \mathcal{L}(\powl)$ be a set of activity labels present in $M$. Let $G = \{A_1, \dots, A_n\}$ be a partitioning of $A$ into $n\geq 2$ parts. We say that $G$ is a conflict group in $M$ if and only if for all $A_i, A_j \in G$:
\[
A_i \neq A_j \implies \forall_{a_i \in A_i, a_j \in A_j} \ a_i\#_{M}a_j. 
\]
We use $\#(M)$ to denote the set of all conflict groups in $M$. 
\end{definition}

\emph{Maximal conflict groups} are those conflict groups not contained within any larger ones, i.e., the set of maximal conflict groups in $M$ is: 
\[
\#^{max}(M) = \{G \in \#(M) \mid \nexists_{G' \in \#(M)} \ G \subset G' \}.
\]

\subsection{Mining for Loops}\label{sec:loop_identification}
To discover loop constructs, our algorithm needs to identifies instances of submodels that are semantically identical within the partial orders. Based on the equivalence relation (cf. \autoref{def:equivalent}), the set of all distinct POWL models in a multiset of partial orders can be partitioned into \emph{POWL equivalent classes}. Each class groups together all models that are semantically identical.

\begin{definition}[POWL Equivalence Classes]\label{def:powl_equivalent_classes}
Let $V$ be a set of POWL models. The equivalence classes of $V$ is a partitioning of V defined as follows:
\[
{\eq}(V) \ = \{\{\powl'\in V \mid \powl' \eq \powl\} \mid \powl \in V\}.
\]
\end{definition}

If an equivalence class in ${\eq}(\mathcal{V}(M))$ contains more than one model, it signifies that this particular model structure is repeated within $M$. All instances within such an equivalence class are then abstracted into a single Loop model, where one instance forms the do-part of the loop.

\subsection{Identifying Coupled Components} \label{sec:cooccurrence_grouping}
In many processes, certain sets of activities or sub-processes exhibit strong synchronous behavior: they tend to appear together or are entirely absent together across all observed instances. Such groups represent logically cohesive blocks of work. Identifying and abstracting these \emph{co-occurring groups} significantly enhances the accuracy of other pattern detection steps (e.g., it enables the discovery of loops at a higher level of abstraction). 

The \emph{node co-occurrence relation} captures whether two nodes always appear together within the same partial order.

\begin{definition}[Node Co-Occurrence Relation]\label{def:cooccur}
Let $M$ be a multiset of partial orders. Let $\powl, \powl' \in \mathcal{V}(M)$ be two nodes in $M$. We say that $\powl$ and $\powl'$ co-occur in $M$, denoted by $\powl \leftrightarrow_{M} \powl'$, if and only if:
\[\forall_{(V, \po) \in M} \ \powl \in V \Leftrightarrow \powl' \in V.\]
\end{definition}

We use the node co-occurrence relation to partition the set of all nodes into disjoint subsets, where all nodes within a particular subset always co-occur.

\begin{definition}[Node Co-Occurrence Partitioning]\label{def:cooccur_part}
Let $M$ be a multiset of partial orders. The co-occurrence partitioning of the nodes of $M$ is defined as follows: 
\[
{\leftrightarrow}(M)\ = \{\{\powl' \in \mathcal{V}(M) \mid \powl' \leftrightarrow_M \powl\} \mid \powl \in \mathcal{V}(M)\}.
\]
\end{definition}

\subsection{Combining Partial Orders} \label{sec:order_aggregation}
After abstracting control-flow patterns, the remaining nodes (which can be atomic transitions or more complex POWL submodels) are aggregated into a single, encompassing partial order that represents the overall process behavior.

\begin{definition}[Combining Partial Orders]\label{def:combine_orders_skips_formal}
Let $M$ be a multiset of partial orders.
The function $\mathbf{CombineOrders(M)}$ returns a single aggregated partial order $(V_{agg}, \po_{agg})$ where $V_{agg} = \mathcal{V}(M)$ and $\po_{agg}$ is constructed as follows:

\begin{itemize}

    \item \textbf{Base Precedence Relation:}
    Let $\po_{base} \subseteq V_{agg} \times V_{agg}$ be the relation defined as follows for any $u, v \in V_{agg}$:
    \[
    u \po_{base} v \iff\]
    \[
     \exists_{(V, \po) \in M} \ u \po v \ \wedge \ \nexists_{(V, \po) \in M} \Big(\{u, v\} \subseteq V \ \wedge \ u \notpo v \Big).
    \]
     This relation captures sequential dependencies that are present and never contradicted in the input partial orders.
    
    \item \textbf{Extended Precedence Relation:}
    Let $\po_{ext} \subseteq V_{agg} \times V_{agg}$ be the relation defined as follows for any $u, v \in V_{agg}$:
    \[
    u \po_{ext} v \iff
    \]
    \[u \closure{\po_{base}} v \ \land \ \nexists_{(V, \po) \in M} \Big(\{u, v\} \subseteq V \ \wedge \ u \notpo v \Big).
    \]
     This relation cautiously extends $\po_{base}$ by inferring additional dependencies by transitivity unless they contradict the input partial orders.
    
    \item \textbf{Transitive Aggregated Relation:}
    $\po_{agg}\ = \mathit{Prune}(\po_{ext})$, where $\mathit{Prune}$ is function that iteratively identifies and resolves violations of transitivity to ensure $\po_{agg}$ is a partial order. For instance, if $v_1 \po_{ext} v_2$ and $v_2 \po_{ext} v_3$ but $v_1 \notpo_{ext} v_3$, an edge (i.e., $(v_1, v_2)$ or $(v_2, v_3)$) is removed. This pruning process is applied iteratively, until no such patterns remain and transitivity is reached. The choice of which edge to remove and the removal order if multiple violations exist will lead to non-deterministic results. 
\end{itemize}

\end{definition}

\subsection{The Recursive POWL Discovery Algorithm}

\begin{algorithm}[!t]
\caption{POWL Discovery from Partial Orders}
\scriptsize
\label{alg:discover_powl}
\DontPrintSemicolon
\KwIn{A non-empty multiset of partial orders $M$}
\KwOut{A single POWL model}

\SetKwFunction{FDiscoverPOWL}{DiscoverPOWL}
\SetKwFunction{FIdentifyXOR}{MineXORs}
\SetKwFunction{Cluster}{MineCooccurringGroups}
\SetKwFunction{FAggregatePO}{CombineOrders}
\SetKwProg{Fn}{Function}{:}{}
\DontPrintSemicolon

\newcommand{\Comment}{}

\nonl\Fn{\FDiscoverPOWL{$M$}}{

    \SetKwBlock{Step}{Step 1: XOR Mining}{}
    \nonl\Step{
        \For{$G = \{A_1, \dots, A_n\} \in \#^{max}(M)\label{line:xor}$}{ 
            \For{$1\leq i \leq n$}{
                $V_i \leftarrow \{\powl \in \mathcal{V}(M) \mid \mathcal{L}(\powl) \subseteq A_i\}$\;
                $\powl_i \leftarrow \FDiscoverPOWL(M\proj_{V_i})$\;
            }
            $\powl_{G} \leftarrow \xor(\powl_1, \dots, \powl_n)$\;
            $M \leftarrow M_{V_1 \cup \dots \cup V_n \rightarrow \powl_{G}}$\; 
        }
    }
    
    \SetKwBlock{Step}{Step 2: Co-Occurrence Grouping}{}
    \nonl\Step{
        \If{$\card{{\leftrightarrow}(M)} > 1$\label{line:cluster}}{
            \For{$V \in {\leftrightarrow}(M)$}{
                
                $\powl_{V} \leftarrow \FDiscoverPOWL(M\proj_{V})$\; 
                $M \leftarrow M_{V \rightarrow \powl_{V}}$\;    
            }
        }
    }
    
    \SetKwBlock{Step}{Step 3: Loop Mining}{}
    \nonl\Step{
        \For{$V \in {\eq}(\mathcal{V}(M))\label{line:loop}$}{
            \If{$\card{V} > 1$}{
                $\powl \leftarrow random(V)$\;
                $\powl_{\Loop} \leftarrow \Loop(\powl, (\tau, 1))$\;
                $M \leftarrow M_{V \rightarrow \powl_{\Loop}}$\;
            }
        }
    }
    
    \SetKwBlock{Step}{Step 4: Skip Mining}{}
    \nonl\Step{
        \For{$\powl \in \mathcal{V}(M)\label{line:skips}$}{
            \If{$\exists_{(V, \po) \in M} \ \powl \notin V$}{
                $\powl_{\xor} \leftarrow \xor(\powl, (\tau, 1))$\;
                $M \leftarrow M_{\{\powl\} \rightarrow \powl_{\xor}}$\;
            } 
        }
    }
    
    \SetKwBlock{Step}{Step 5: Order Aggregation}{}
    \nonl\Step{
        $\powl = (V, 
        \po) \leftarrow \FAggregatePO(M)$\; 
        \KwRet{\powl}
    }
}
\end{algorithm}

With the foundational operations and pattern definitions in place, we now present the recursive algorithm for discovering a POWL model from a multiset of POTs, outlined in \autoref{alg:discover_powl}. Steps 1-3 identify and abstract common process patterns (choices, co-occurring groups, and loops), aiming for a more structured and compact model. The core steps 4-5 handle optionality and aggregate remaining elements into a single partial order.

\begin{enumerate}

\item \textbf{XOR Mining:} The algorithm identifies maximal conflict groups $A_1, \dots, A_n$. For each $A_i$, the corresponding set of nodes $V_i$ is determined, and a submodel $\powl_i$ is recursively discovered from the projection $M\proj_{V_i}$. All branch models are combined with the operator $\xor$.

\item \textbf{Co-Occurrence Grouping:} The algorithm builds the co-occurrence partitioning ${\leftrightarrow}(M)$. For each group $V$, it recursively discovers a POWL model $\powl_V$ from the projection $M\proj_{V}$, and it then replaces $V$ with $\powl_V$.

\item \textbf{Loop Mining:} The algorithm partitions $\mathcal{V}(M)$ into POWL equivalent classes ${\eq}(\mathcal{V}(M))$. If a class $V$ contains multiple instances, a representative model $\powl$ is chosen, and a loop $\Loop(\powl, (\tau, 1))$ is created to replace $V$.

\item \textbf{Skip Mining:} For each node $\powl \in \mathcal{V}(M)$, if it's found to be absent in at least one partial order in $M$, it's replaced with $\xor(\powl, (\tau,1))$.

\item \textbf{Partial Order Aggregation:}
The algorithm synthesizes a single partial order from the nodes in $M$, as detailed in \autoref{sec:order_aggregation}.

\end{enumerate}

\subsection{Discussion: Soundness and Fitness Guarantees}\label{sec:discussion}
POWL models are inherently \emph{sound} by construction \cite{DBLP:journals/is/KouraniZSA25}. Beside soundness, our algorithm provides a \emph{perfect fitness} guarantee. In this context, perfect fitness means that any sequence of activities that is a valid linearization of any POT in the input multiset can be replayed by the discovered POWL model. This guarantee primarily stems from (i) the conservative order aggregation (Step 5), which avoids inferring precedences conflicting with input POTs, and (ii) from handling optionality (Step 4), allowing replication of traces with missing parts. These mechanisms, along with the abstraction steps (Steps 1-3), which are also fitness-preserving, ensure that all observed behaviors in the input POTs are permissible in the final POWL model.


\section{Implementation and Evaluation}\label{sec:example}

We have implemented our discovery approach in the Python library `powl`\footnote{Install via `pip install powl`. A usage example is available at \url{https://github.com/humam-kourani/POWL/blob/main/examples/partial_order_based_discovery.py}.}. and developed a web application for demonstration purposes\footnote{The demonstrator is accessible at \url{https://po-aware-powl-miner.streamlit.app/}.}.

To evaluate our approach, we conducted an experiment on two well-known, real-life event logs: BPI Challenge 2012 and BPI Challenge 2017 \footnote{Both event logs are available at \url{https://data.4tu.nl/}}. We created several variants of each log by filtering them to retain only the top 4, 6, 8, and 12 most frequent activities, in addition to using the full, unfiltered logs.

We compare our approach against two state-of-the-art techniques: the Zebra Miner \cite{kovavr2024exploratory} and the eST\textsuperscript{2} Miner \cite{folz2025est}\footnote{Zebra Miner is accessible at \url{https://www.fernuni-hagen.de/ilovepetrinets/zebra/}, while eST\textsuperscript{2} Miner is available in ProM (\url{https://promtools.org/}).}. To derive the input partial orders for the eST\textsuperscript{2} Miner, we used the `PTrace' plugin in ProM with the `timePO' option selected. 
To ensure a fair comparison, we configured the eST\textsuperscript{2} Miner with a fitness threshold of 1.0 to produce perfectly fitting models. For other parameters, we used the same defaults selected in \cite{folz2025est}. 

A timeout of one hour was set for each discovery run. The quality of the discovered models was measured using standard alignment-based \emph{fitness} and \emph{precision}, computed with PM4Py\footnote{\url{https://processintelligence.solutions/pm4py}.}. As standard conformance checking algorithms operate on totally ordered traces, these metrics were computed using only the `complete' instances from the event logs. These derived sequential traces represent valid linearizations of the more general partially ordered traces.



\begin{table}[!t]
\centering
\caption{Evaluation Results. T/O indicates reaching a timeout of one hour.}
\label{tab:ev_res}
\resizebox{\textwidth}{!}{%
\begin{tabular}{|l|l|>{\centering\arraybackslash}p{1.2cm}|>{\centering\arraybackslash}p{1.2cm}|>{\centering\arraybackslash}p{1.2cm}|>{\centering\arraybackslash}p{1.2cm}|c|>{\centering\arraybackslash}p{1.2cm}|>{\centering\arraybackslash}p{1.2cm}|>{\centering\arraybackslash}p{1.2cm}|>{\centering\arraybackslash}p{1.2cm}|c|}
\hline
\multicolumn{2}{|l|}{Event Log} &  \multicolumn{5}{c|}{BPI Challenge 2012} & \multicolumn{5}{c|}{BPI Challenge 2017}\\
\hline  
\multicolumn{2}{|l|}{Num. Activities}  & 4 & 6 & 8  & 12   & Full (24) & 4 & 6    & 8    & 12   & Full (26) \\
\hline
Time (sec) & POWL Miner & 8 & 12      & 13 & 19   & 25  & 17      & 26   & 24   & 33   & 41  \\
     & eST\textsuperscript{2} Miner  & 145 & 799 & 2683 & T/O  & T/O & 897 & T/O  & T/O  & T/O  & T/O \\
     \hline
Precision  & POWL Miner & 0.53    & 0.51    & 0.56     & 0.45 &  0.21   & 0.95    & 0.71 & 0.56 & 0.53 & 0.27      \\
     & eST\textsuperscript{2} Miner  & 0.45    & 0.39    & 0.46     & -  & - & 0.91    & -  & -  & -  & -    \\
     \hline
\end{tabular}%
}
\end{table}

\paragraph{Results}
All models generated during the experiment are available at \url{https://github.com/humam-kourani/results-po-aware-powl-miner}. The results of our experiment are summarized in \autoref{tab:ev_res}. The results for the Zebra Miner are excluded as it reached the time limit on all log variants. We don't report fitness values as both the POWL Miner and eST\textsuperscript{2} Miner achieved perfect fitness on all logs due to their fitness guarantees. 

\begin{figure}[!t]
    \centering

        \includegraphics[width=\linewidth]{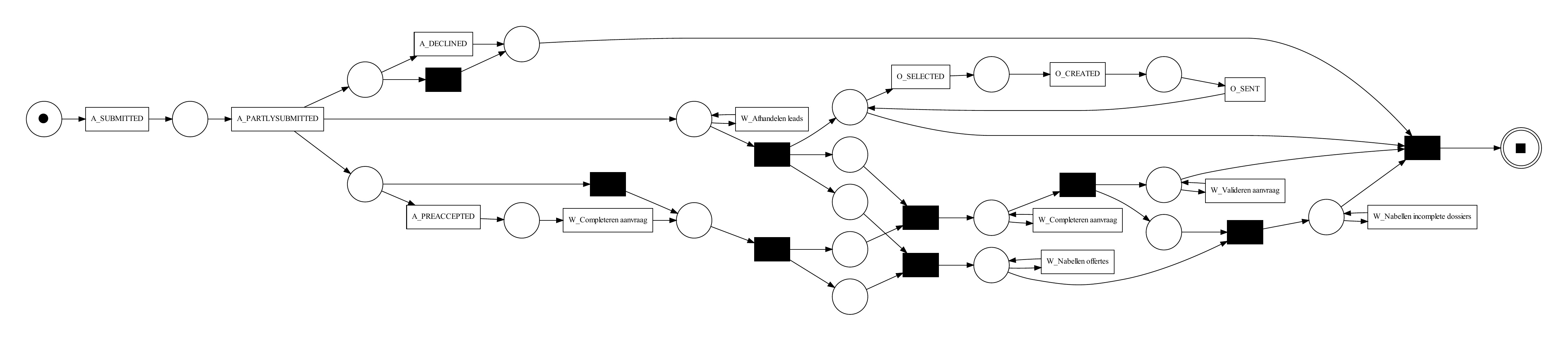}

    \caption{The POWL model discovered from the BPI Challenge 2012 event log with 12 activities translated into a Petri net.}
    \label{fig:bpic12_models}
\end{figure}

In terms of precision, our approach consistently outperformed the eST\textsuperscript{2} Miner. This suggests that our methodology produces a more faithful and less over-generalized representation of the process. For example, \autoref{fig:bpic12_models} shows the model discovered for the BPIC 2012 log with 12 activities. Regarding performance, our algorithm demonstrates high scalability. It successfully processed all log variants, with the longest discovery task completing in just 42 seconds for the full BPIC 2017 log. In contrast, the eST\textsuperscript{2} Miner was significantly slower and encountered the one-hour timeout on the BPIC 2012 log with 12 activities and already on the BPIC 2017 log with just 6 activities, failing to produce a model for the more complex scenarios.

\section{Conclusion and Future Directions}\label{sec:conclusion}
This paper presented a novel process discovery algorithm that leverages the Partially Ordered Workflow Language (POWL) to discover process models from partially ordered event data. Recognizing the limitations of traditional techniques that often linearize concurrent behavior, our approach leverages the partial order nature of events by first transforming process traces into partial orders. Building upon this foundation, a hierarchical mining algorithm systematically combines these partial orders while identifying and abstracting choice (XOR) and loop constructs. The resulting models are inherently sound and achieve perfect fitness with respect to the input partial orders. We demonstrated the scalability of our approach by applying it on complex real-life event data.

An important avenue for future work is to investigate incorporating noise filtering mechanisms to abstract from infrequent or exceptional behaviors. We believe this research contributes to advancing process discovery by providing a more faithful and understandable representation of operational processes where activities frequently overlap or run in parallel.

\textbf{Acknowledgments.} 
This work was funded by the Federal Ministry of Research, Technology and Space (BMFTR), Germany (grant 01IS23065).


\bibliographystyle{splncs04}
\bibliography{references_compact}

\end{document}